\documentclass[aps,prl,twocolumn,superscriptaddress]{revtex4}
\usepackage{amsmath}
\usepackage{graphicx}
\usepackage{multirow}
\usepackage{subfigure}
\usepackage{color,soul}

\newcommand{\ket}[1]{|#1\rangle}
\newcommand{\bra}[1]{\langle#1|}

\newcommand{\eq}{\begin{equation}}
\newcommand{\fine}{\end{equation}}

\begin{document}
\title{Experimental optimal cloning of four-dimensional quantum states of photons}

\author{E. Nagali}

\affiliation{Dipartimento di Fisica dell'Universit\`{a} ``La
Sapienza'', Roma 00185, Italy}

\author{D. Giovannini}

\affiliation{Dipartimento di Fisica dell'Universit\`{a} ``La
Sapienza'', Roma 00185, Italy}

\author{L. Marrucci}

\affiliation{Dipartimento di Scienze Fisiche, Universit\`{a} di
Napoli ``Federico II'', Compl.\ Univ.\ di Monte S. Angelo, 80126
Napoli, Italy}

\affiliation{CNR-SPIN, Compl.\ Univ.\ di Monte S. Angelo, 80126
Napoli, Italy}

\author{S. Slussarenko}

\affiliation{Dipartimento di Scienze Fisiche, Universit\`{a} di
Napoli ``Federico II'', Compl.\ Univ.\ di Monte S. Angelo, 80126
Napoli, Italy}

\author{E. Santamato}

\affiliation{Dipartimento di Scienze Fisiche, Universit\`{a} di
Napoli ``Federico II'', Compl.\ Univ.\ di Monte S. Angelo, 80126
Napoli, Italy}

\author{F. Sciarrino}

\email{fabio.sciarrino@uniroma1.it}

\affiliation{Dipartimento di Fisica dell'Universit\`{a} ``La
Sapienza'', Roma 00185, Italy}

\affiliation{CNR-Istituto Nazionale di Ottica, Firenze, Italy}

\begin{abstract}
Optimal quantum cloning is the process of making one or more copies
of an arbitrary unknown input quantum state with the highest
possible fidelity.
All reported demonstrations of quantum cloning have so far been limited to copying two-dimensional quantum states, or qubits.
We report the experimental realization of the optimal quantum
cloning of four-dimensional quantum states, or ququarts, encoded
in the polarization and orbital angular momentum degrees of freedom
of photons. Our procedure, based on the symmetrization method, is
also shown to be generally applicable to quantum states of
arbitrarily high dimension -- or qudits -- and to be scalable to
an arbitrary number of copies, in all cases remaining optimal.
Furthermore, we report the bosonic coalescence of two single-particle entangled states.
\end{abstract}

\maketitle

Classical information
can be freely measured, perfectly copied  
on demand, and broadcast
without fundamental limitations. The handling of quantum
information, which is encoded in the quantum states of physical
systems, is instead subject to several 
fundamental restrictions. For example, an unknown quantum state of an individual
system cannot be measured completely, unless we have infinite
identical copies at our disposal. For a finite number of copies $N$,
the state estimation can only be partial, and it can be
characterized by an average ``fidelity'' lower than one (where one
corresponds to perfect state identification). It has been proven that
the optimal value of such state estimation fidelity is given by
$F_{est}^{d}(N) = (N + 1)/(N + d)$, where $d$ is the dimension of
the quantum space \cite{Gisi97}. A similar restriction is posed by
the \emph{quantum no-cloning theorem}, stating that an unknown
quantum state cannot be copied perfectly \cite{Woot82}. It is
however possible to make imperfect copies, characterized by a
cloning fidelity lower than one \cite{Scar05}. Starting with $N$
identical copies of the input state and generating 
$M>N$
output optimal copies, the optimal copying fidelity is given by $F_{clon}^{d}(N,M)=\frac{M-N+N(M+d)}{M(N+d)}$, for the
case of ``symmetric'' cloning, that is for a uniform fidelity of all
copies \cite{Nave03}. It is important to note that, for a given
input, the optimal cloning fidelity is always higher than the
corresponding optimal state-estimation fidelity, reducing to the
latter in the limit $M\rightarrow\infty$ \cite{Brus99}. Therefore, the optimal
quantum cloning process is useful whenever one needs to broadcast
quantum information among several parties without measuring
it in the process. Quantum cloning thus represents an important
multipurpose tool of the emerging quantum information technology.
Let us stress that the 
 advantage of
quantum cloning over state estimation grows for an increasing
dimension $d$ of the quantum state. More specific applications of
quantum cloning are found in the security assessment of quantum
cryptography, the realization of minimal disturbance measurements,
the enhancement of the transmission fidelity over a lossy quantum
channel, and the separation of classical and quantum information
\cite{Gisi02,Ricc05}.

It is well known that all tasks of quantum information can be
performed using only two-dimensional quantum states, or qubits.
However, it has been recently recognized that significant
fundamental and practical advantages can be gained by employing
higher dimensional quantum states instead, or qudits. For
example, quantum cryptographic protocols based on qudits may achieve
improved security, entangled qudits can show increased resistance
to noise, a qudit-based quantum computation may require less
resources for its implementation, and the use of quantum computing
as physics simulators can be facilitated by using qudits
\cite{Kasz00,Cerf02,Coll02,Geno05,Lany09,Neel09,Vert10}.


Light quantum states can be used for implementing qudits, either by
exploiting many-photon systems \cite{More06,Vall07,Lany08} or by
combining different degrees of freedom of the same photon (
``hybrid'' states) such as linear momentum, arrival time, and
orbital angular momentum (OAM) or other transverse modes
\cite{Torr03,Vazi03,Moli04,Lang04}. In particular,
we have  recently reported the first experimental generation and tomography of
hybrid qudits with dimension $d=4$, also dubbed ququarts, that
were encoded in the polarization and OAM of single photons
\cite{Naga10}.


In this paper, we report the 
realization of the optimal
quantum cloning $1\rightarrow2$ (i.e., $N=1, M=2$) of 
ququarts ($d=4$)
encoded in the polarization and OAM of single photons. The cloning
process is based on the symmetrization technique
\cite{Ricc04,Scia04,Irvi04}, that has been recently proven
theoretically to be optimal for arbitrary dimension $d$
\cite{Naga09b}. The simultaneous control of polarization and OAM was
made possible thanks to the q-plate, a photonic device introducing a
spin-orbit angular momentum coupling \cite{Marr06,Naga09a}. 

Let us recall the working principle of the symmetrization method for
$1\rightarrow2$ quantum cloning, in the case of a generic
$d$-dimensional quantum state (qudit)
\cite{Ricc04,Scia04,Irvi04,Naga09b}. The input qudit
$\ket{\varphi}_s$ is sent into one input port (mode $\mathbf{k}_s$)
of a balanced beam splitter (BS). In the other BS input port (mode
$\mathbf{k}_a$) we send an ancilla photon in the fully mixed random
$d$-dimensional state
$\rho_a=\frac{I_d}{d}=\frac{1}{d}\sum_n{\ket{n}_a\bra{n}} $. The
basis is here chosen such that $\ket{1}\equiv\ket{\varphi}$. After
the interaction in the BS, we consider only the case of the two
photons emerging in the same output mode. The cloning fidelity is defined as the
average overlap between the quantum state of each output photon
emerging from the BS and the input photon state $\ket{\varphi}$.
Hence, we can distinguish two cases, depending on the ancilla state:
(i) the input state is $\ket{1}_s\ket{1}_a$ or (ii) it is any of
the other $d-1$ states $\ket{1}_s\ket{n}_a$ with $n\neq1$. In the
first case, Hong-Ou-Mandel (HOM) quantum interference due to the
bosonic symmetry leads to a doubled probability of having the two
photons emerging in a common output BS mode, as compared to the
second case. So, once given that the two photons emerge in the same
mode (i.e., successful cloning occurs), then the first case has a
relative probability of $2/(d+1)$ while the second case has a total
probability $(d-1)/(d+1)$. Since the first case corresponds to a
fidelity of 1 (both photons are identical to the input one) and the
second of 0.5 (one photon is identical, the other is orthogonal), we
obtain the following average cloning fidelity $F=\frac{1}{2}+\frac{1}{d+1}$,
that corresponds just to the upper bound for the cloning fidelity
$F_{clon}^{d}(N,M)$ given above, for $N=1$ and $M=2$.
\begin{figure}
\centering
\includegraphics[scale=.30]{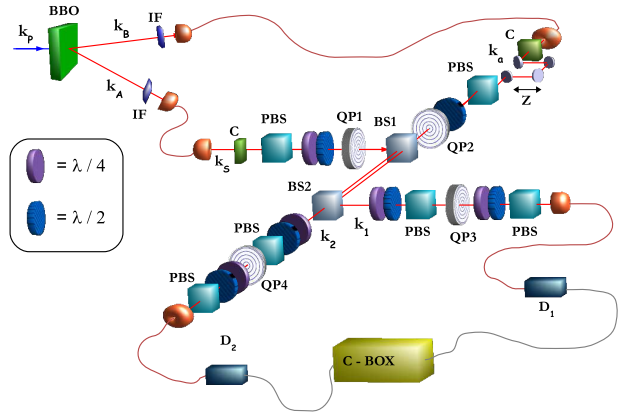}
\caption{Experimental apparatus for implementing the $1\rightarrow
2$ optimal quantum cloning of polarization-OAM photon ququarts. See
main text for a definition of all symbols. \label{setup}}
\end{figure}

This cloning procedure has been previously demonstrated
experimentally only for photonic qubits, either encoded in the
polarization space $\pi$ \cite{Scia04} or in the bidimensional OAM
subspace $o_2$ spanned by $m=\pm2$ \cite{Naga09b}, where $m$ is the
OAM eigenvalue per photon along the beam axis in units of $\hbar$.
We now consider photonic ququarts encoded in the four-dimensional
``spin-orbit'' space $\pi \otimes o_2$, i.e.~obtained as tensor
product of the polarization space and the OAM subspace with
$m=\pm2$. A generic separable state in this space will be indicated
as $\ket{\varphi,\ell}=\ket{\varphi}_{\pi}\ket{\ell}_{o_2}$, where
$\ket{\cdot}_{\pi}$ and $\ket{\cdot}_{o_2}$ stand for the
polarization and OAM quantum states, respectively. We introduce in
this space a first basis $\{\ket{1_\text{I}},\ket{2_\text{I}},\ket{3_\text{I}},\ket{4_\text{I}}\}$,
hereafter called ``logic'' basis, corresponding to
$\{\ket{R,+2},\ket{R,-2},\ket{L,+2},\ket{L,-2}\}$, where $R$ ($L$)
refers to right (left) circular polarization and the $\pm2$ integers
refer to the OAM eigenvalue. Given this first basis, we may introduce
four other bases such that they are all \emph{mutually unbiased} to
each other \cite{Klap04}. In our spin-orbit space $\pi \otimes o_2$,
all the states belonging to the five mutually unbiased basis can be
generated and detected by an appropriate combination of birefringent
wave plates, polarizing beam-splitters, q-plates and single mode
fibers, as 
described in Ref.\ \cite{Naga10}. Three of these
bases (bases I, II, and III) are formed of separable states of
polarization and OAM  of the photon (e.g., basis I is the logical one defined above),
while the remaining two (bases IV and V) are formed of entangled
states of these two degrees of freedom (see
\cite{Naga10} for the complete list of states). In particular, the
IV basis, which will be utilized in the present work, is composed of
the following four states: $(\ket{R,+2}\pm\ket{L,-2})/\sqrt{2}$ and
$(\ket{L,+2}\pm\ket{R,-2})/\sqrt{2}$.

We now consider the optimal quantum cloning of a photonic ququart in
the spin-orbit space $\pi \otimes o_2$. Being $d=4$, we expect an optimal cloning fidelity $F=7/10$. If
$\ket{\varphi}=\ket{1}$ is the input state, each output cloned
photon (tracing out the state of the other photon) is hence expected
to be found in the mixed state:
\begin{equation}
\rho_1=\rho_2=\frac{1}{10}(7\ket{1}\bra{1}+\ket{2}\bra{2}+\ket{3}\bra{3}+\ket{4}\bra{4}),
\label{eqrho}
\end{equation}
where the set $\{\ket{1},\ket{2},\ket{3},\ket{4}\}$ forms a basis in
the space $\pi \otimes o_2$ (not necessarily the logical one). In
particular, we will experimentally test the outcome of the quantum
cloning procedure for all the four states of the logical basis I and
for all the four states of the IV basis, corresponding to entangled
spin-orbit states.

The experimental layout is schematically reported in
Fig.~\ref{setup}. A $\beta$-barium borate crystal (BBO) cut for
type-II phase matching, pumped by the second harmonic of a
Ti:Sapphire mode-locked laser beam, generates via spontaneous
parametric fluorescence photon pairs on modes $k_A$ and $k_B$ with
linear polarization, wavelength $\lambda=795$ nm, and pulse
bandwidth $\Delta\lambda=4.5$ nm, as determined by two interference
filters (IF). The coincidence rate of the source is equal to 18 kHz.
Photons generated on mode $k_A$ and $k_B$ are delivered to the setup
via single mode fibers, thus defining their transverse spatial modes
to a pure TEM$_{00}$, corresponding to OAM $m=0$. After the fiber
output, two wave plates (C) compensate the polarization rotation
introduced by the fibers and a polarizing beam-splitter (PBS)
projects the polarization on the horizontal state $\ket{H}_{\pi}$.
Then on mode $k_s$ the ququart to be cloned is encoded in the single
photon polarization and OAM through a ququart preparation stage,
based on a combination of wave plates, a q-plate, a PBS (only for
bases I, II, III), and additional wave plates (see Ref.\
\cite{Naga10} for details). On mode $k_a$, for quantum cloning the
ancilla photon is prepared in a fully mixed state
$\rho_a=\frac{I_{\pi}}{2}\frac{I_{o2}}{2}$, i.e. fully randomized
both in polarization and in OAM. This is obtained by randomly
rotating, during each experimental run, a half-wave plate inserted
before the q-plate QP2 and by randomly inserting or removing
another half-wave plate located after the same q-plate. The time
delay between photons on mode $k_s$ and $k_a$ was set to zero by an
adjustable delay line (Z), in order to ensure the interference
condition necessary for the optimal quantum cloning process within
the balanced beam splitter BS1. A second beam splitter (BS2) is then
used to separate the two photons emerging from the same output port
of BS1, allowing post-selection of this outcome by coincidence
detection. On both output modes $k_1$ and $k_2$ of BS2 we perform a
full ququart state measurement, by combining a standard polarization
analysis set and an OAM analysis set, the latter based on the
quantum transferrer $o_2\rightarrow\pi$ \cite{Naga09a}. Depending on
the specific ququart basis being used, the detailed setting of this
ququart measurement stage varies slightly, as discussed in
\cite{Naga10}. Finally, the output photons are coupled into single
mode fibers and detected by single-photon counters (D$_1$ and D$_2$)
connected to the coincidence box (C-BOX) recording the
time-coincident photon detections.

\begin{figure}[h]
\textbf{(a)}
\begin{minipage}[m]{\linewidth}
\centering
\includegraphics[width=0.6\linewidth,bb=20 20 760 550,clip]{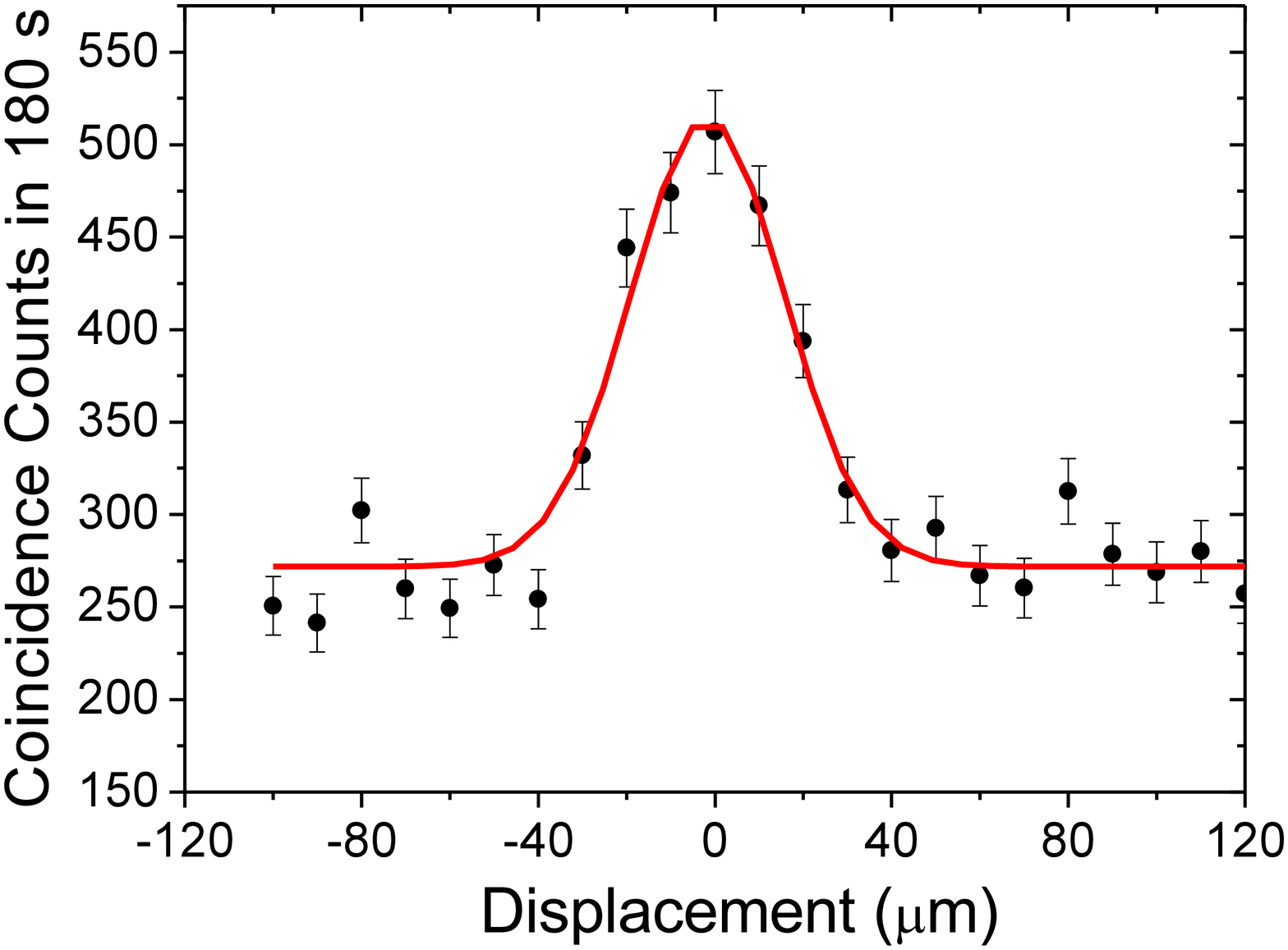}
\end{minipage}
\begin{minipage}[m]{0.58\linewidth}
\centering
\textbf{(b)}
\includegraphics[width=0.85\linewidth,bb=75 20 315 255,clip]{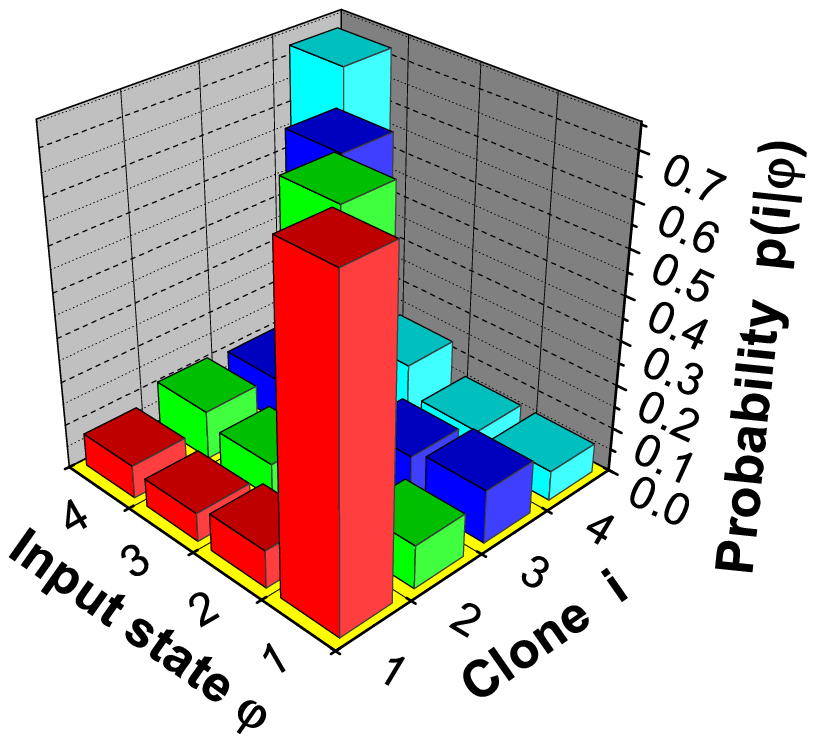}
\end{minipage}
\begin{minipage}[m]{0.40\linewidth}
\textbf{(c)}
\centering
\begin{tabular}{c|c}
\multicolumn{1}{c|}{\textbf{State}} &
\multicolumn{1}{c}{\textbf{Fidelity}} \\
\hline
$\ket{1_\text{I}}$ & $(0.740\pm0.016)$ \\
$\ket{2_\text{I}}$ & $(0.677\pm0.012)$ \\
$\ket{3_\text{I}}$ & $(0.707\pm0.012)$ \\
$\ket{4_\text{I}}$ & $(0.708\pm0.017)$ \\
\end{tabular}
\end{minipage}
\caption{Experimental results of the cloning process for the ququart
states belonging to the I basis 
\textbf{a)} Hong-Ou-Mandel coalescence for two input photons
prepared in state $\ket{L,-2}$. The
enhancement peak is of $R=(1.90\pm0.13)$. \textbf{b)} Probability
$p(i|\varphi)$ of detecting a clone in any output state $\ket{i}$ of
the basis, for any given input state $\ket{\varphi}$ of the same
basis. \textbf{c)} Experimental cloning fidelities for the four
input states.} \label{base1}
\end{figure}

As a first step, we have verified the occurrence of the HOM
interference between the two photon ququarts impinging on modes
$k_s$ and $k_a$ of BS1.
The ancillary photon was prepared in the same quantum state as the signal photon, in order for the interference to occur.
The two-photon coincidence counts were measured as a function of the
optical path delay between $k_s$ and $k_a$. In
Fig.~\ref{base1}-\textbf{a} we report an example of the results we
obtained for the case of an input state belonging to the logical
basis. The HOM peak is observed 
with a measured
coincidence enhancement $R=(1.89\pm0.05)$, consistent with the
theoretical value $R_{th}=2$. Once ensured a good interference
condition between the two photons, we moved on to testing the
quantum cloning for each of the four input states of the logical
basis. 
Being
$\ket{\varphi}$ the input state to be cloned, the measurement stage
on mode $k_1$ has been set so as to filter only outcoming photons in
state $\ket{\varphi}$ 
while on mode $k_2$ all four possible outcomes
$\ket{i}$ of the logical basis have been detected. We have thus
recorded the corresponding coincidence counts $N_{\varphi,i}$. These
coincidence counts give us an estimate of the probability
$p(i|\varphi)$ of each photon clone to be found in any specific
state $\ket{i}$ of the basis, with $i=1,2,3,4$, as a function of the
input state $\ket{\varphi}$, regardless of the state of the other
photon. In particular, we have
$p(i|\varphi)=N_{\varphi,i}/N$ for $i\neq\varphi$ and
$p(\varphi|\varphi)=(N_{\varphi,\varphi}+\sum_{i\neq\varphi}N_{\varphi,i})/N$,
where $N=N_{\varphi,\varphi}+2\sum_{i\neq\varphi}N_{\varphi,i}$. The
factor 2 appearing in the expression of $N$ takes into account the
additional coincidences that would be detected by swapping the
measurements performed on modes $k_1$ and $k_2$, for $i\neq\varphi$,
that of course are equal to $N_{\varphi,i}$ in the average. The
theoretical
values for these probabilities are given by
the corresponding coefficients in the clone density matrix in Eq.\
(\ref{eqrho}). The cloning fidelity is $F=p(\varphi|\varphi)$. The
experimental results obtained when cloning all states of the logical
basis are reported in Fig.~\ref{base1}-\textbf{b,c}. The measured
values of the fidelity, as well as their average value
$\overline{F}_I=(0.708\pm0.007)$, are all in good agreement with the
theoretical prediction $F=0.7$.

\begin{figure}[h]
\textbf{(a)}
\begin{minipage}[m]{\linewidth}
\centering
\includegraphics[width=0.6\linewidth,bb=20 20 760 550,clip]{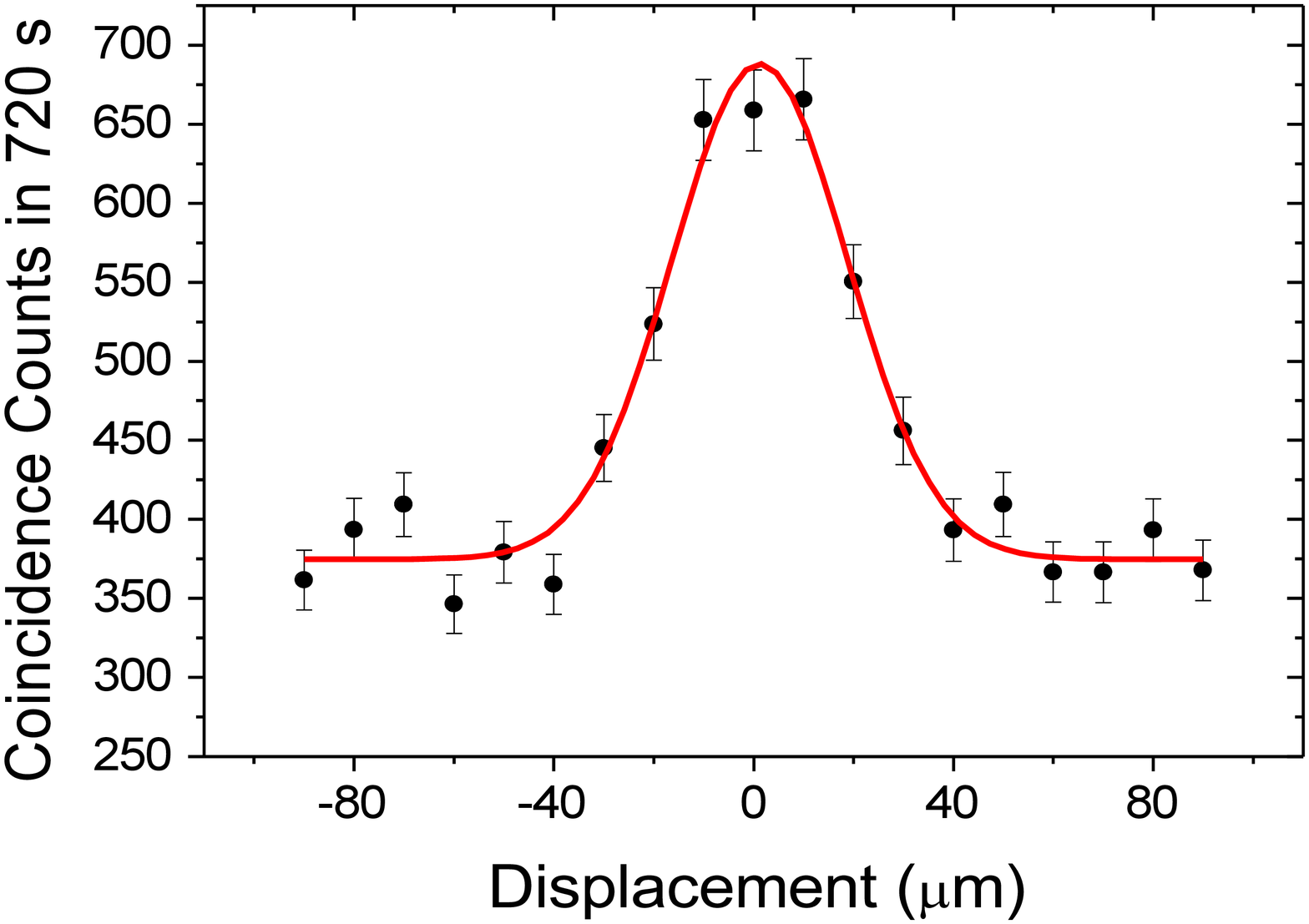}
\end{minipage}
\begin{minipage}[m]{0.58\linewidth}
\centering
\textbf{(b)}
\includegraphics[width=0.85\linewidth,bb=75 20 315 255,clip]{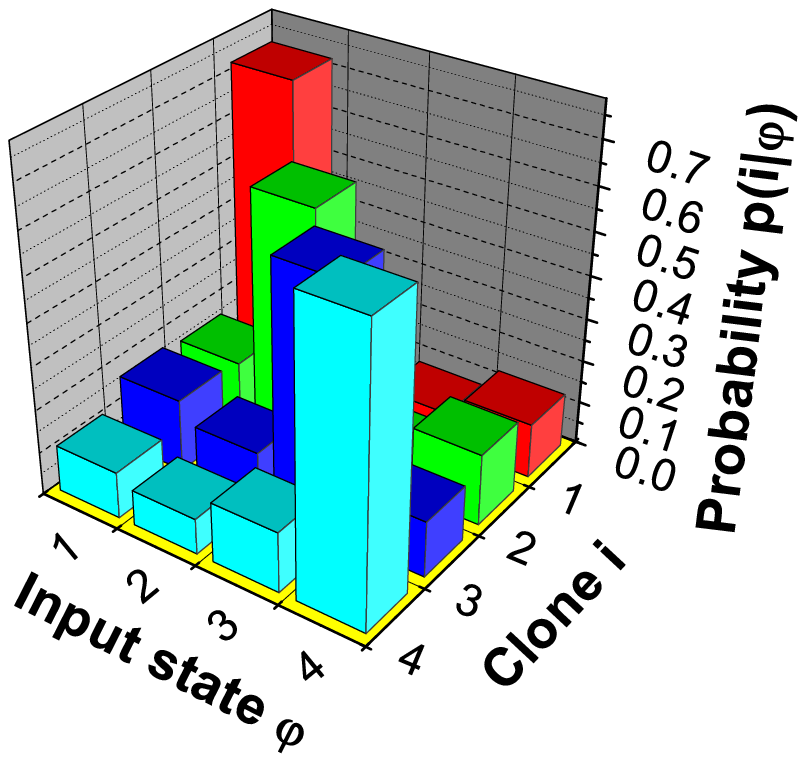}
\end{minipage}
\begin{minipage}[m]{0.40\linewidth}
\textbf{(c)}
\centering\begin{tabular}{c|c}
\multicolumn{1}{c|}{\textbf{State}} &
\multicolumn{1}{c}{\textbf{Fidelity}} \\
\hline
$\ket{1_\text{IV}}$ & $(0.726\pm0.010)$ \\
$\ket{2_\text{IV}}$ & $(0.582\pm0.004)$ \\
$\ket{3_\text{IV}}$ & $(0.580\pm0.005)$ \\
$\ket{4_\text{IV}}$ & $(0.662\pm0.008)$ \\
\end{tabular}
\end{minipage}
\caption{Experimental results of the cloning process for all ququart
states belonging to the IV basis, made of entangled spin-orbit
states. \textbf{a)} Hong-Ou-Mandel coalescence for two input photons
prepared in state $2^{-1/2}(\ket{R,+2} + \ket{L,-2})$. The
enhancement peak is of $R=(1.84\pm0.05)$. \textbf{b)} Probability $p(i|\varphi)$ of detecting a
clone in an output state $\ket{i}$ of the IV basis, for any given
input state $\ket{\varphi}$ of the same IV basis.\textbf{c)} Experimental cloning fidelities for all four input states.} \label{base4}
\end{figure}

Cloning only states belonging to the logical basis is clearly not
enough to demonstrate the generality of our cloning procedure.
We have therefore repeated the cloning
experiment for all ququart states belonging to the IV basis, that
includes spin-orbit entangled states.
These states can be generated and analyzed by exploiting the q-plate
capability of entangling and disentagling the polarization and the
OAM degree of freedom of a photon. This required removing the first
PBS on modes $k_1$ and $k_2$ and properly setting the orientation of
all wave plates \cite{Naga10}. In Fig.~\ref{base4} we report the
experimental results of the HOM peak for one of the states and of
the cloning of all states belonging to the IV basis. Fig.\ref{base4}-a, in particular, demonstrates interference between two single-particle entangled states.
As expected, this measurement underlines how the bosonic coalescence of two particles is not tied to the indistinguishability of each individual degree of freedom, but rather that of the whole quantum state (whether each of such states is entangled, as is the case in our experiment, or separable).
As can be inferred from the table in Fig.~\ref{base4}c, the cloning fidelities
are again in reasonable agreement with the expected one, and the
average fidelity value reads $\overline{F}_{IV}=(0.638\pm 0.004)$.
The small discrepancy with respect to the theoretical expectations
is quantitatively well explained by the imperfect randomization of
the ancilla photon (which 
is found to be somewhat
unbalanced), the slightly lower HOM enhancement achieved ($R=1.84$),
and the non-unitary preparation and analysis fidelities ($\sim0.9$) \cite{FidEnt}.
We stress that the setup alignment has not been
re-optimized for cloning states of basis IV, in order to
properly test the universality of our cloning apparatus. We notice that
the average value of the quantum cloning fidelity is much larger than the one expected
for the quantum state estimation on a single copy, equal to $0.4$ \cite{Brus99}.

\begin{figure}[t!!]
\centering
\includegraphics[scale=0.35,bb=0 0 550 150,clip]{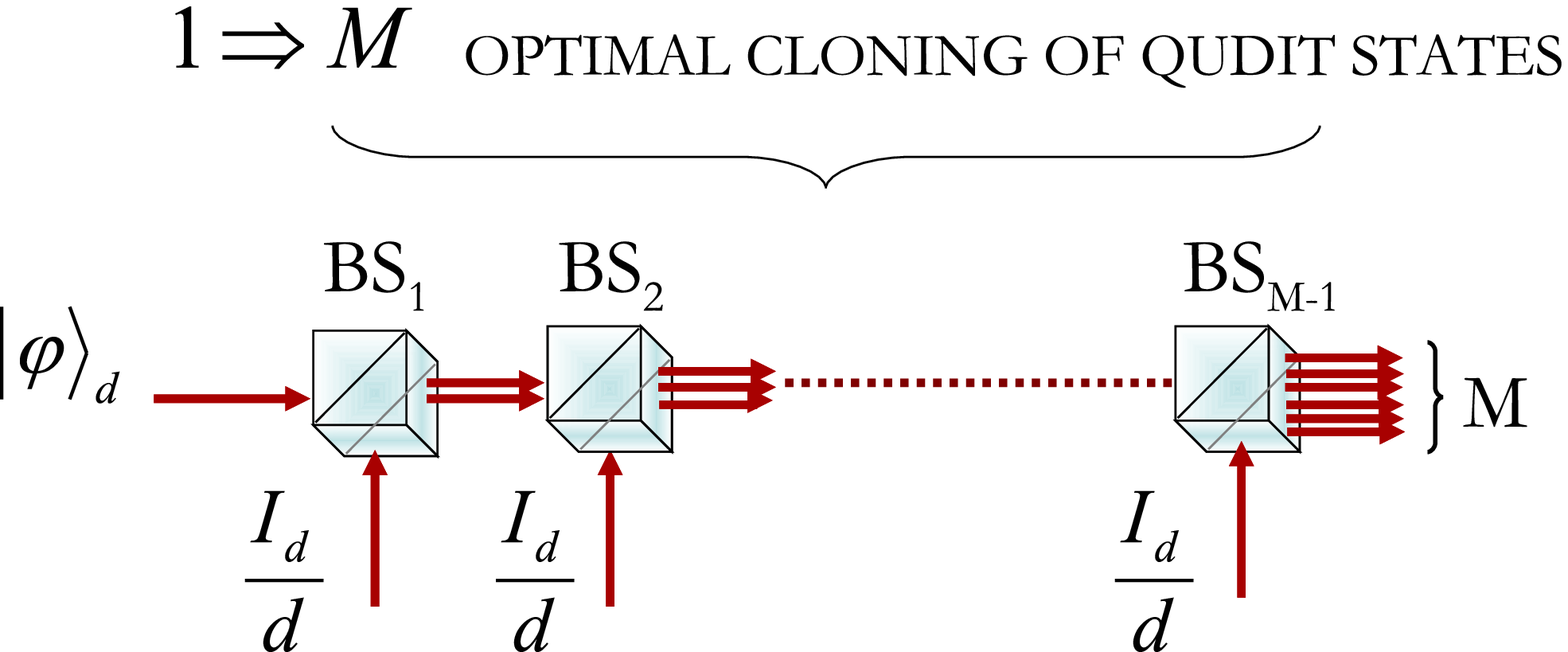}
\caption{Schematic representation of the $1\rightarrow M$ quantum
cloning process of a qudit state by cascading the symmetrization
technique.} \label{logic}
\end{figure}

Finally, we note that the symmetrization procedure for the cloning
of photonic qudits, that we have experimentally demonstrated here
for the $1\rightarrow2$ case, can be 
scaled up to the general
$N\rightarrow M$ cloning \cite{Masu05}, that is starting with $N$ identical input
photons and generating $M>N$ copies. The main idea, illustrated in Fig.~\ref{logic},
 is that of using a cascaded configuration of $M-N$
beam splitters and $M-N$ ancilla photons in fully mixed states: Fig.4. A more exhaustive demonstration of this result will
be presented in a forthcoming paper.

In summary, we have implemented the optimal quantum cloning
$1\rightarrow 2$ of ququart states encoded in the polarization and
OAM degrees of freedom of a single photon.
 This work was supported by project HYTEQ - FIRB, Finanziamento Ateneo 2009 of Sapienza
Universit\`{a} di Roma, and european project PHORBITECH of the FET program (grant 255914).


\begin{thebibliography}{99}
\bibitem{FidEnt} The weights of the four components of the ancilla mixed state in the basis IV have been experimentally found to be: $w_1 = 0.32$, $w_2 = 0.16$, $w_3 = 0.16$, and $w_4 = 0.36$. By considering the preparation and analysis  fidelity and the factor $R$, we computed the expected fidelity values $F_1 = 0.702$, $F_2 =0.594$, $F_3 =0.597$, and $F_4 =0.68$, with mean value $F_{IV} = 0.645$, in good agreement with the experimental results.
\end{thebibliography}
\end{document}